\documentclass[aps,twocolumn,prd,showpacs,showkeys,preprintnumbers,superscriptaddress,nobibnotes,floatfix,longbibliography,nofootinbib]{revtex4-1}
\pdfoutput=1
\usepackage{amsmath}
\usepackage{amsfonts}
\usepackage{amssymb}
\usepackage{mathrsfs}
\usepackage{graphicx}
\usepackage{color}
\usepackage{longtable}
\usepackage{hyperref}
\usepackage{multirow}
\usepackage{slashed}
\usepackage{epsf}

\usepackage{color}

\begin{document}

\title{
Probing the Supersymmetric Grand Unified Theories at the Future Proton-Proton Colliders and Hyper-Kamiokande Experiment
}

\author{Waqas Ahmed}
\email{waqasmit@nankai.edu.cn}

\affiliation{${}^{a}$ School of Physics, Nankai University, No.94 Weijin Road, Nankai District, Tianjin, China}
\author{Tianjun Li}
\email{tli@mail.itp.ac.cn}
\affiliation{ CAS Key Laboratory of Theoretical Physics, Institute of Theoretical Physics, Chinese Academy of Sciences, Beijing 100190, China}
\affiliation{School of Physical Sciences, University of Chinese Academy of Sciences, No. 19A Yuquan Road, Beijing 100049, China}
\author{Shabbar Raza}
\email{shabbar.raza@fuuast.edu.pk}
\affiliation{Department of Physics, Federal Urdu University of Arts, Science and Technology, Karachi 75300, Pakistan}
\author{Fang-Zhou Xu}
\email{xfz14@mails.tsinghua.edu.cn} 
\affiliation{Institute of Modern Physics, Tsinghua University, Beijing 100084, China}


\begin{abstract}

Gauge coupling unification in the Supersymmetric Standard Models strongly implies the
Grand Unified Theories (GUTs). With the grand desert hypothesis, we show that 
the supersymmetric GUTs can be probed at the future proton-proton (pp) colliders and 
Hyper-Kamiokande experiment. For the GUTs with the GUT scale $M_{GUT} \le 1.0\times 10^{16}$~GeV,
we can probe the dimension-six proton decay via heavy gauge boson exchange 
at the Hyper-Kamiokande experiment. Moreover, for the GUTs with 
$M_{GUT} \ge 1.0\times 10^{16}$~GeV, we for the first time study the upper bounds 
on the gaugino and sfermion masses. We show that the GUTs with anomaly and 
gauge mediated supersymmetry breakings are well within the reaches of 
the future 100 TeV pp colliders such as the ${\rm FCC}_{\rm hh}$ and SppC,
and the supersymmetric GUTs with gravity mediated supersymmetry breaking 
can be probed at the future 160 TeV pp collider. 

\end{abstract}
\maketitle


\textbf{Introduction.--}Supersymmetry (SUSY) provides a natural solution to the gauge hierarchy problem 
in the Standard Model (SM).  In the supersymmetric SMs (SSMs) with R-parity, 
gauge coupling unification can be achieved~\cite{gaugeunification}, 
the Lightest Supersymmetric Particle (LSP) 
such as the lightest neutralino can be a dark matter (DM) candidate~\cite{Jungman:1995df}, 
and  the  electroweak  (EW)  gauge  
symmetry  can  be  broken  radiatively  due  to  the  large  top quark Yukawa coupling, etc.  
In particular, gauge coupling unification strongly suggests 
Grand Unified Theories (GUTs)~\cite{Georgi:1974sy,Pati:1974yy,Mohapatra:1974hk,Fritzsch:1974nn,Georgi:1974my}, 
which may be constructed from superstring theory.  Therefore, supersymmetry is a bridge between 
the low energy phenomenology and high-energy fundamental physics, and thus is 
the promising new physics beyond the SM.

However, after the LHC Run 2, the null results of the SUSY searches have given strong constraints 
on the SSMs. For example, the low mass bounds on the gluino, first-two generation squarks,
stop, and sbottom are about 2.3~TeV, 1.9~TeV, 1.25~TeV, and 1.5~TeV, 
respectively~\cite{ATLAS-SUSY-Search, Aad:2020sgw, Aad:2019pfy, CMS-SUSY-Search-I, CMS-SUSY-Search-II}.
Thus, there might exist SUSY EW fine-tuning (EWFT) problem. And there are some promising and successful solutions available 
in literatures, for example, Refs.~\cite{Dimopoulos:1995mi, Cohen:1996vb, Kitano:2005wc, Kitano:2006gv, LeCompte:2011cn,
LeCompte:2011fh, Fan:2011yu, Kribs:2012gx, Baer:2012mv, Baer:2012cf, 
Drees:2015aeo,Ding:2015epa,Baer:2015rja,Batell:2015fma,Fan:2014axa}. 
In particular, in the Super-Natural SUSY~\cite{Leggett:2014hha,Du:2015una, Li:2015dil}, it was shown that 
the fine-tuning measure defined by Ellis-Enqvist-Nanopoulos-Zwirner~\cite{Ellis:1986yg} 
and Barbieri-Giudice~\cite{Barbieri:1987fn}
is at the order of one naturally, despite having relatively heavy supersymmetric particle (sparticle) spectra.
The previous natural SSMs generically predict some relatively light sparticles, for instance, Higgsino, stop, gluino,
 and sleptons, which can be tested at the future proton-proton (pp) colliders such as the 
${\rm FCC}_{\rm hh}$~\cite{Benedikt:2018csr} and 
SppC~\cite{CEPC-SPPCStudyGroup:2015csa}.

Because the gauge coupling unification in the SSMs strongly suggests GUTs, the
interesting and challenging question is: can we probe the supersymmetric GUTs at the future pp colliders 
and other experiments even if there does exist the SUSY EWFT problem? 
If yes, what is the center-of-mass energy of the future pp collider needed? We shall study it in this paper. 
In the GUTs, the well-know prediction is the dimension-six proton decay $p\to e^+ \pi^0$ via heavy 
gauge boson exchange, and the proton lifetime is given by~\cite{Dutta:2016jqn}
\begin{eqnarray}
\tau_p(e^+\pi^0) &\simeq& 1.0\times 10^{34} \times
\left(\frac{2.5}{A_R}\right)^2 \times
\left(\frac{0.04}{\alpha_{\rm GUT}}\right)^2
\nonumber \\  &&
\times
\left(\frac{M_{\rm GUT}}{1.0\times 10^{16}~{\rm GeV}}\right)^4
~{\rm years} ~,~\,
\label{eq:proton}
\end{eqnarray}
where $A_R$ is  the dimensionless one-loop renormalization factor associated with anomalous dimension 
of the relevant baryon-number violating operators, $\alpha_{\rm GUT}$ is the unified gauge coupling,
and $M_{\rm GUT}$ is the GUT scale. The current lower limit on the proton lifetime from
the Super-Kamiokande experiment is $\tau_p > 1.6 \times 10^{34}$ years~\cite{Miura:2016krn}.
Thus, we obtain $M_{\rm GUT} \ge 1.0\times 10^{16}~{\rm GeV}$.
At the future Hyper-Kamiokande experiment, we can probe the proton lifetime
at least above $1.0 \times 10^{35}$ years~\cite{Abe:2018uyc}.
Therefore, the GUTs with $M_{\rm GUT} \le 1.0\times 10^{16}~{\rm GeV}$ is within the reach of 
the future Hyper-Kamiokande experiment.

In the following, with the grand desert hypothesis from the EW scale to the GUT scale,
 we shall show that the supersymmetric GUTs with $M_{\rm GUT} \ge 1.0\times 10^{16}~{\rm GeV}$ 
can be probed at the future pp colliders. 
The supersymmetry searches at the 100~TeV pp colliders have been 
studied previously~\cite{Benedikt:2018csr, Cohen:2013xda, Arkani-Hamed:2015vfh,
Fan:2017rse, Golling:2016gvc}. For the integrated luminosity 30~${\rm ab}^{-1}$,
Wino via Bino decay, gluino ${\tilde g}$ via heavy flavor decay, gluino via light flavor decay,
first-two generation squarks ${\tilde q}$, and stop can be discovered for their masses
up to about 6.5~TeV, 11~TeV, 17~TeV, 14~TeV, and 11~TeV, respectively.
Moreover, if the gluino and first-two generation squark masses are similar,
they can be probed up to 20 TeV.

Moreover, in the SSMs, supersymmetry is broken in the hidden sector, and then
 supersymmetry breaking is mediated to the SM observable sector via gravity
mediation~\cite{chams, bbo, cmssm}, 
gauge mediation~\cite{Dine:1993yw, Giudice:1998bp, Meade:2008wd}, 
or anomaly mediation~\cite{Randall:1998uk, Giudice:1998xp}.
For the supersymmetric GUTs with $M_{GUT} \ge 1.0\times 10^{16}$~GeV, 
we for the first time study the upper bounds on the gaugino and sfermion masses.
We show that the GUTs with anomaly and gauge mediated supersymmetry
breakings are well within the reaches of the future 100 TeV pp colliders 
such as the ${\rm FCC}_{\rm hh}$ and SppC,
and the supersymmetric GUTs with gravity mediated supersymmetry breaking can be probed 
at the future 160 TeV pp collider. 
The interesting viable parameter spaces for gravity mediation,
which can be probed at the ${\rm FCC}_{\rm hh}$ and SppC, have been discussed as well.

%

\begin{figure}[t!]
\includegraphics[width=1.0\columnwidth]{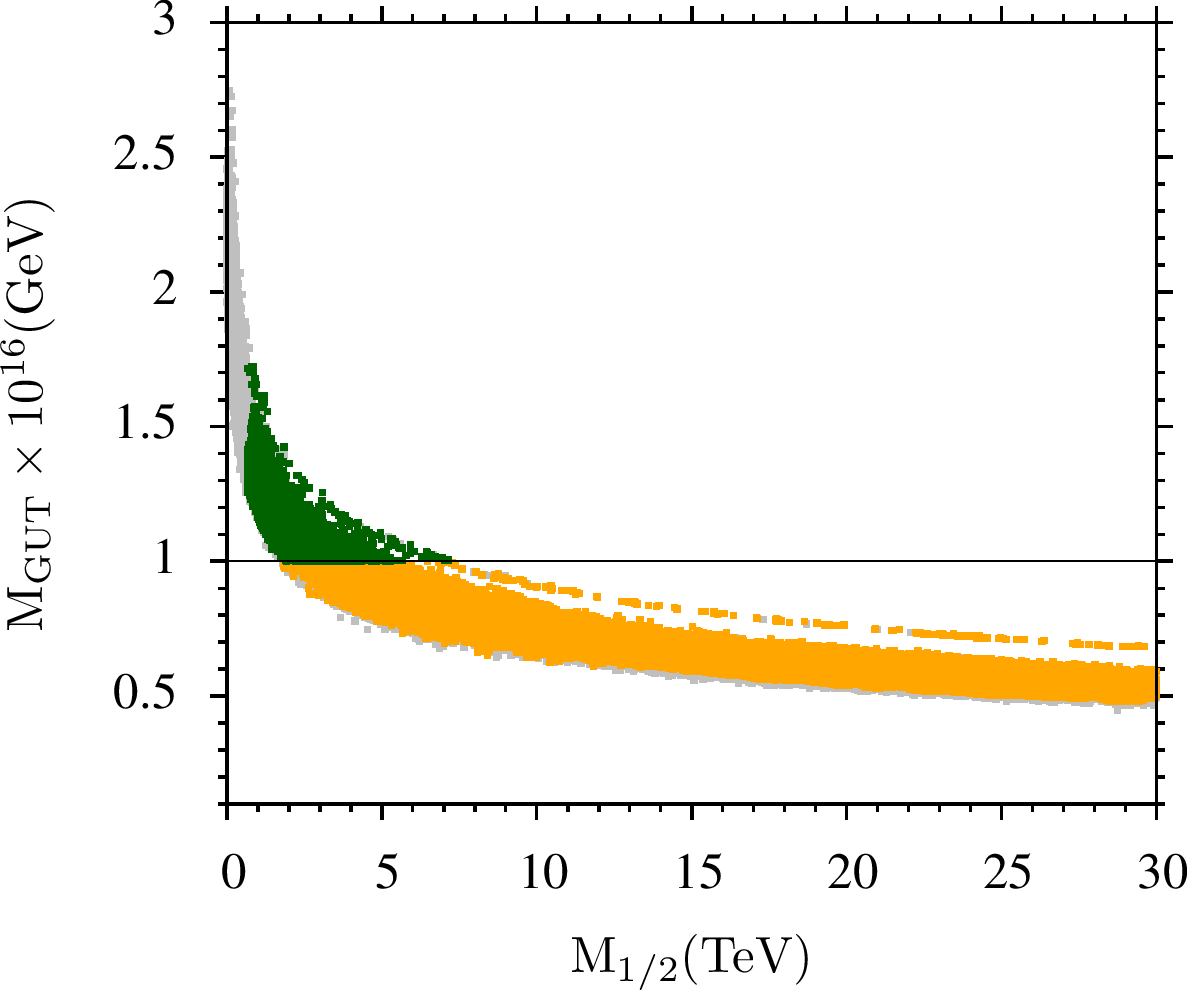}
\caption{
Gray points are consistent with the REWSB and LSP neutralino. Orange points satisfy the mass bounds 
 and the constraints from rare $B-$meson decays. Green points are a subset of orange points 
and satisfy $M_{GUT}\gtrsim 1\times 10^{16} \,{\rm GeV}$.  
}
\label{fig:MUm12}
\end{figure}

%

%
\textbf{Scanning Codes and Constraints.--}We use the ISAJET~7.85 package~\cite{ISAJET} to 
perform random scans over the parameter space of gravity mediated SUSY breaking via
the  minimal supergravity (mSUGRA)~\cite{chams,bbo} or Constrained MSSM (CMSSM)~\cite{cmssm},
 as well as the anomaly mediated SUSY breaking~\cite{Randall:1998uk,Giudice:1998xp}.
 To study the gauge mediated SUSY breaking~\cite{Dine:1993yw, Giudice:1998bp, Meade:2008wd},
we also  employ the SPheno 4.0.4 package \cite{Porod:2003um} generated with SARAH 4.14.3 \cite{Staub:2008uz}. 

 The collected data points all satisfy the requirement of the Radiative Electroweak Symmetry Breaking (REWSB), 
has the lightest neutralino being the LSP for gravity and anomaly mediations, SM-like Higgs boson mass
$m_h \subset [123,~127]~{\rm GeV}$, and gluino mass $m_{\tilde{g}}\geq 2.2~{\rm TeV}$. 
After collecting the data, we impose
 the constraints from rare decay processes $B_{s}\rightarrow \mu^{+}\mu^{-} $~\cite{Aaij:2012nna}, 
$b\rightarrow s \gamma$~\cite{Amhis:2012bh}, and $B_{u}\rightarrow \tau\nu_{\tau}$ \cite{Asner:2010qj}. 
To be general, we do not require the relic abundance of the LSP neutralino to satisfy the Planck bound within 
$5\sigma$ $0.114 \leq \Omega_{\rm CDM}h^2 (\rm Planck) \leq 0.126$~\cite{Akrami:2018vks}.

\begin{figure*}[ht]
    \centering
        \begin{tabular}{c c}
    \includegraphics[width = 0.5\textwidth]{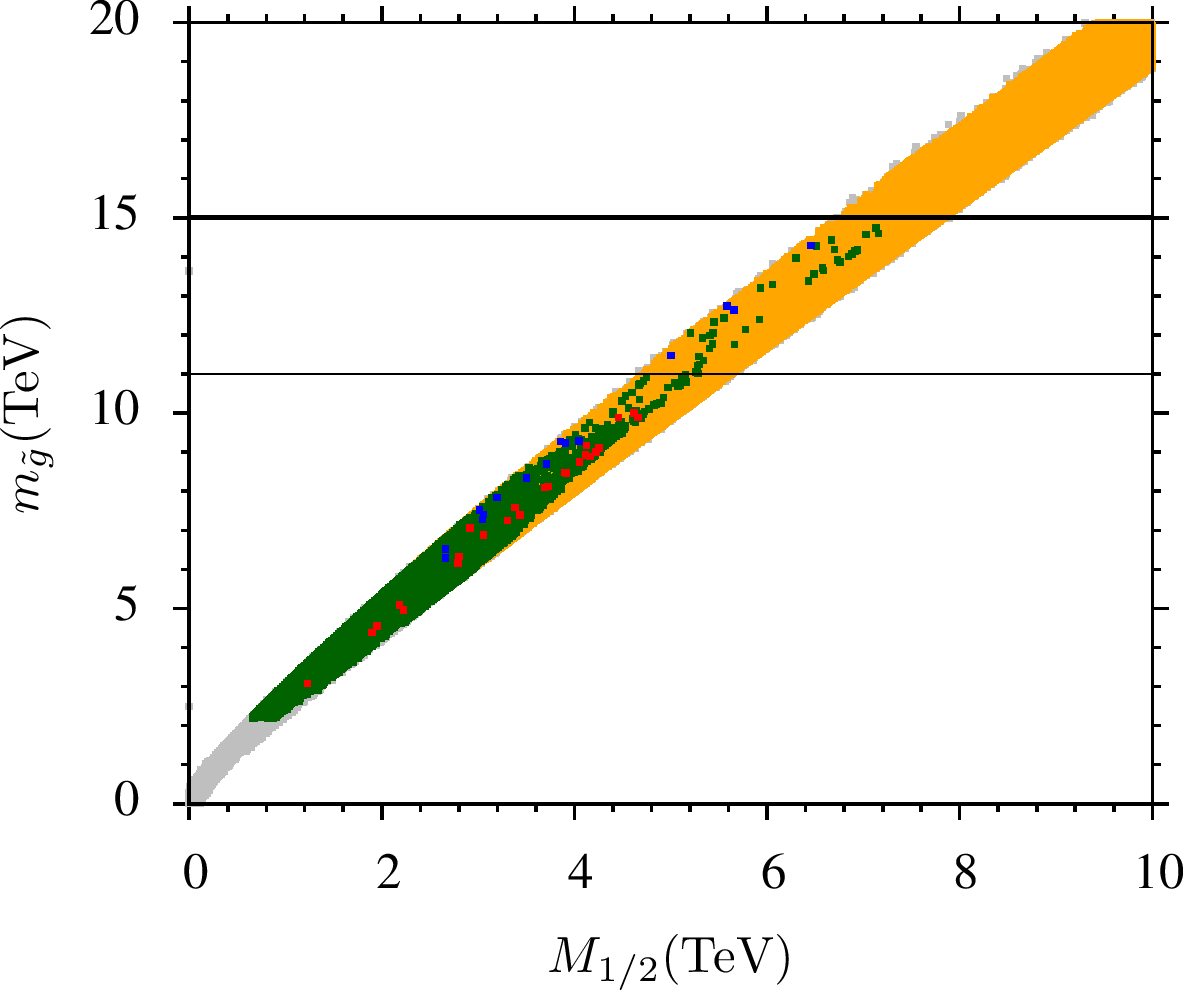}  
 \hspace{-.01cm}
\includegraphics[width = 0.5\textwidth]{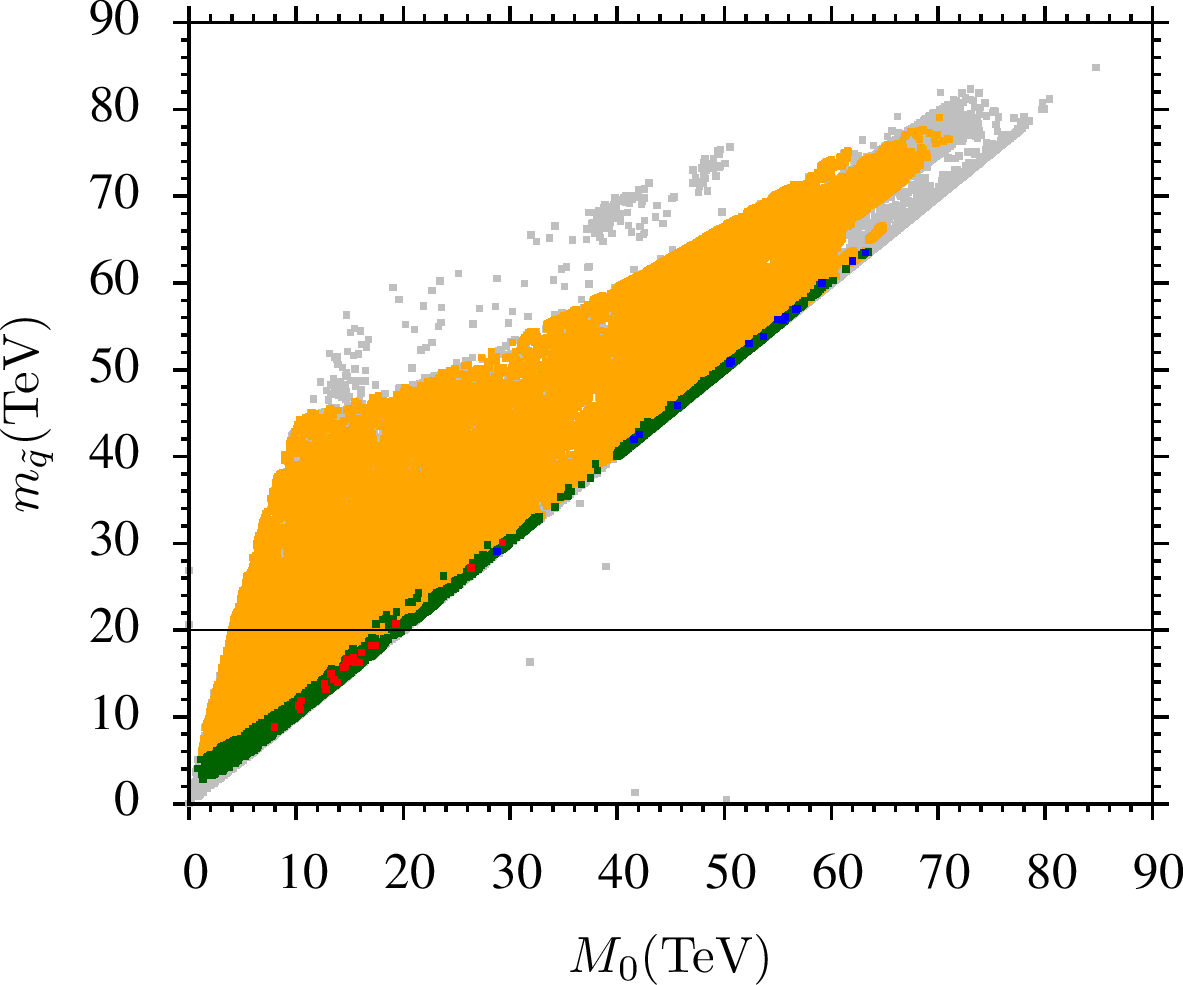} 

\\
    \end{tabular}
    \caption{The color coding for gray, orange, and green points is the same as the Fig.~\ref{fig:MUm12}.
\textbf{Left:} plot in the $m_{\tilde g}$ and $M_{1/2}$ plane.  
Red ($\tan\beta> \, 7.5$) and blue ($\tan\beta < \, 7.5$) points are 
subset of green points and represent solutions which satisfy the Planck 5$\sigma$ bound.
\textbf{Right:} plot in the first two generation squark mass $m_{\tilde q}$ 
and $M_{0}$ plane. 
Red ($\tan\beta> \, 9$) and blue ($\tan\beta < \, 9$) points are 
subset of green points and represent solutions which satisfy the Planck 5$\sigma$ bound.
}
    \label{fig:M12}
\end{figure*}





\textbf{Gravity Mediated Supersymmetry Breaking: mSUGRA/CMSSM.--}The mSUGRA/CMSSM \cite{chams,bbo,cmssm} 
is based on the GUTs and
$N=1$ supergravity where supersymmetry breaking is communicated through the supergravity interaction. 
It is one of the most widely studied SUSY scenarios,  and 
has three supersymmetry breaking soft terms at the GUT scale: the universal gaugino mass $M_{1/2}$, 
universal scalar mass $M_{0}$, and universal trilinear coupling $A_{0}$. The other free parameter 
 $\tan\beta$ is the ratio of vacuum expectation values (VEVs) of two Higgs-doublets,
 and a discrete parameter ${\rm sign}(\mu)=\pm1$. 
We  perform the random scans for the following mSUGRA/CMSSM parameter space
\begin{align}
0\leq  & \, M_{0} \,  \leq  90\, \rm{TeV}, \nonumber \\
0\leq  & \, M_{1/2}    \leq  30\, \rm{TeV}, \nonumber  \\
-3 \leq & \, {A_{0}/M_{0}}     \leq  3, \nonumber \\ 
2 \leq  & \, \tan\beta \, \leq  60
\label{parameterRange}
\end{align}
with $\mu >0$ and $m_t = 173.2\, {\rm GeV}$ \cite{:2009ec}. The results are not too 
sensitive to one or two sigma variations in the value of $m_t$ \cite{bartol2}. 
We use $m_b^{\overline{DR}}(M_{\rm Z})=2.83$ GeV as well  which is hard-coded into the ISAJET.

Because the sfermions in the SSMs form the complete GUT multiplets while
gauginos do not, the universal guagino mass $M_{1/2}$ has big effects on gauge coupling unification.
We present the plot $M_{GUT}$ vs $M_{1/2}$ in Fig.~\ref{fig:MUm12}. In our figures,
gray points are consistent with the REWSB and LSP neutralino. Orange points satisfy the mass bounds 
 and the constraints from rare $B-$meson decays. Green points are a subset of orange points 
and satisfy $M_{GUT}\gtrsim 1\times 10^{16} \,{\rm GeV}$.  
Thus, we obtain that the upper bound 
on $M_{1/2}$ is about 7 TeV. This bound can be translated into the upper bound 15 TeV on gluino mass,
as shown below.

In the left panel of Fig.~\ref{fig:M12}, we show results of our scans 
in $M_{1/2}-m_{\tilde g}$ plane. We first find that the upper bound on the gluino mass is
15 TeV. In addition, the red points ($\tan\beta> \, 7.5$) 
and blue points ($\tan\beta < \, 7.5$) are the subsets of green points and satisfy the
Planck 2018 5$\sigma$ bounds on dark matter relic density. Interestingly, glunio masses for
the red points are lighter than 11 TeV, and thus the glunio for the red points 
is within the reach of the ${\rm FCC}_{\rm hh}$ and SppC~\cite{Cohen:2013xda,Fan:2017rse}.

In the right panel of Fig.~\ref{fig:M12}, we present the scan results in 
the first-two generation squark mass $m_{\tilde q}$ and $M_{0}$ plane. In particular, 
 $M_0$ can be very heavy up to 65 TeV.
Similarly, the red points ($\tan\beta> \, 9$) 
and blue points (($\tan\beta < \, 9$)) are also the subsets of green points and satisfy the
Planck 2018 5$\sigma$ bounds on dark matter relic density.
We see that the maximum value of $M_{0}$ for most of red points is about 20 TeV. 
Because $m^{2}_{\tilde q} \simeq M^{2}_{0}+ (5-6)M^{2}_{1/2}$~\cite{Baer:2006rs} and
the maximum value of $M_{1/2}\sim 7 \,{\rm TeV}$,  we obtain that
the maximum value of the first-two generation squark masses for most of red points is  
about $m_{\tilde q}\simeq 20 \,{\rm TeV}$,  as shown clearly in $m_{\tilde q}-M_{0}$ plot. Thus,
most of the red points can be probed at the ${\rm FCC}_{\rm hh}$ and SppC~\cite{Cohen:2013xda, Golling:2016gvc} 

Because $M_0$ can be very large up to 65 TeV, it will be difficult to search
for the squarks and sleptons at the ${\rm FCC}_{\rm hh}$ and SppC in general. 
Thus, we can look for the gauginos at the future pp colliders.
For the integrated luminosity 30~${\rm ab}^{-1}$ at the ${\rm FCC}_{\rm hh}$ and SppC, 
 gluino via heavy and light flavor decays
 can be discovered for the masses
up to about 11~TeV and 17~TeV, respectively.
Thus, if gluino decays via light flavor squarks, it can be discovered at the ${\rm FCC}_{\rm hh}$ and SppC.
However, in our viable parameter space, the lightest squark is generically to be light stop, and
thus we  do have gluino via heavy flavor decay. To probe such gluino with mass up to 15 TeV,
we find that the center-of-mass energy of the future pp collider needs to be about 160 TeV.
And we can discover Wino at this energy as well.


\begin{figure*}[ht]
    \centering
        \begin{tabular}{c c}
    \includegraphics[width = 0.5\textwidth]{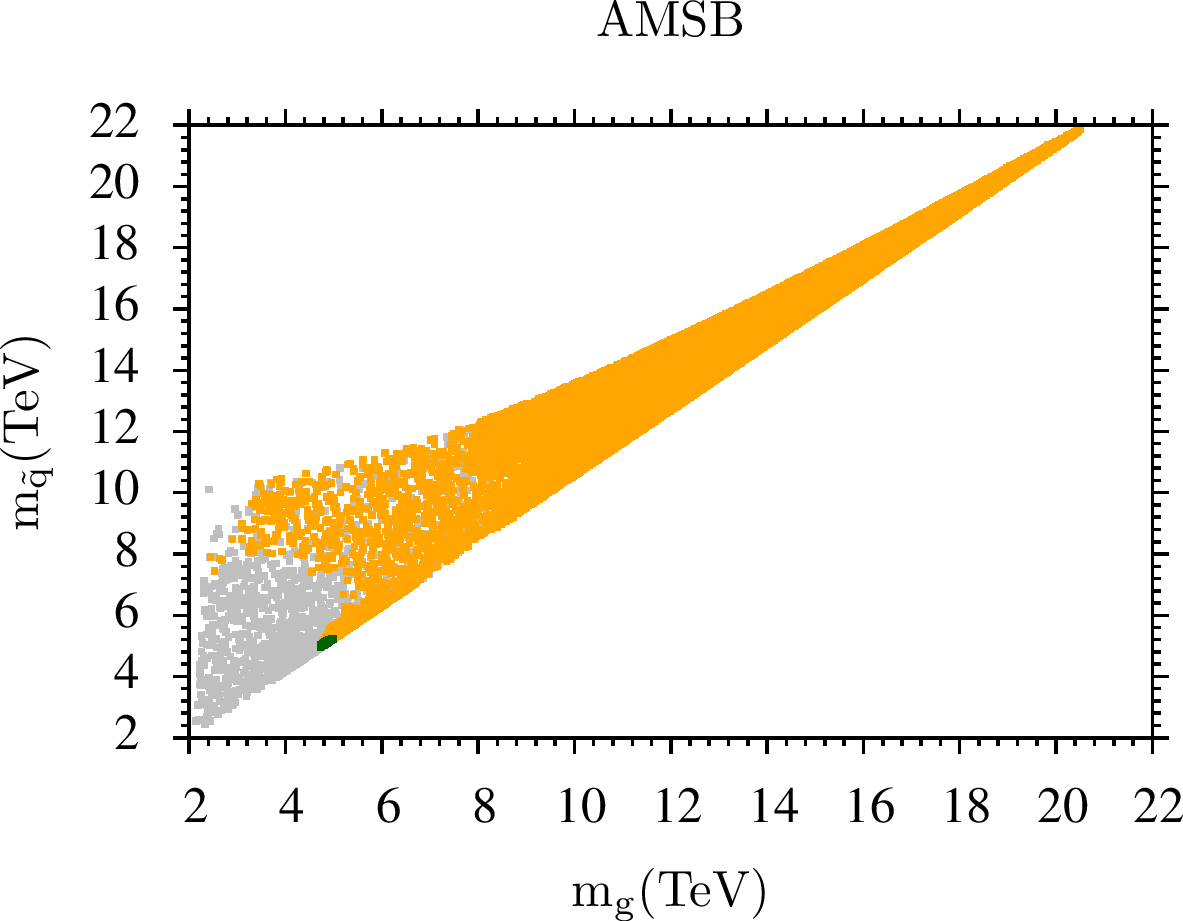} &
        \hspace{-.01cm}\includegraphics[width = 0.5\textwidth]{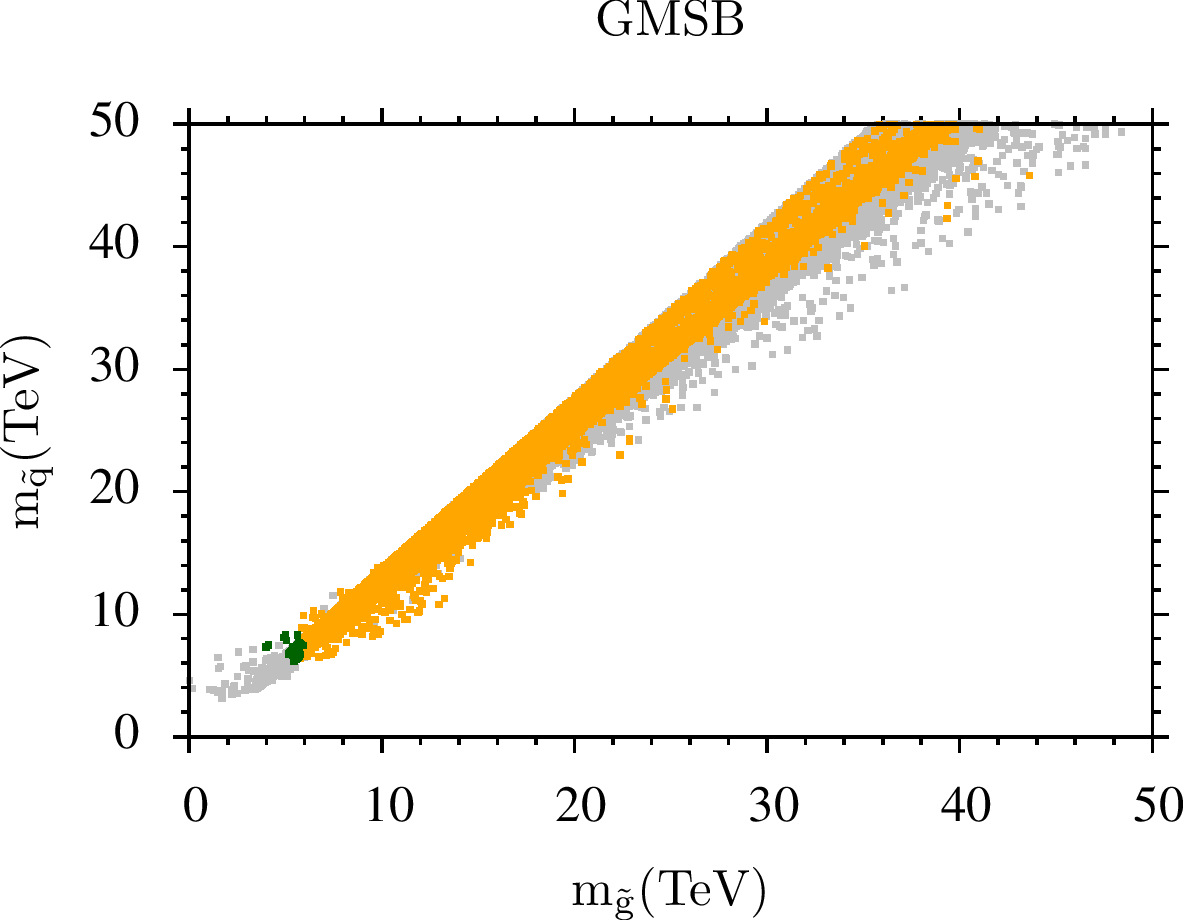}  \\
    \end{tabular}
    \caption{Plots in the gluino mass and first-two generation squarks mass plane.  \textbf{Left} and \textbf{right} panels
are for the minimal AMSB scenario and minimal GMSB scenario, respectively. 
Color coding is same as the Fig.~\ref{fig:MUm12}. }
    \label{fig:GMSB}
\end{figure*}

\textbf{Anomaly Mediated Supersymmetry Breaking.--}Anomaly mediated supersymmetry breaking (AMSB) 
is a special type of gravity mediated SUSY breaking. In this case, SUSY breaking is communicated to 
the visible sector from the hidden sector via a super-Weyl anomaly~\cite{Randall:1998uk,Giudice:1998xp}.  
In the minimal AMSB, there are three basic parameters in addition to ${\rm sign}(\mu)$:  $\tan\beta$, 
the universal scalar mass $M_0$ at the GUT scale which is introduced to solve the tachyonic slepton mass problem,
 and  gravitino mass $M_{3/2}$.
 We have performed the random scans over the following parameter space of the minimal AMSB
\begin{align}
1 \, \rm{TeV}\ ,\leq  & \, M_{0} \,  \leq  75\, \rm{TeV}, \nonumber \\
100\,\rm{TeV} \, \leq  & \, M_{3/2}    \leq  30\, \rm{TeV}, \nonumber  \\
2 \leq  & \, \tan\beta \, \leq  60
\label{parameterRange-AMSB}
\end{align}
with $\mu >0$ and $m_t = 173.2\, {\rm GeV}$ \cite{:2009ec}. In the left panel of Fig.~\ref{fig:GMSB}, 
we present the results of our scan in the $m_{\tilde q}-m_{\tilde g}$ plane. 
All the points, which satisfy the current experimental constraints and have
 $M_{U} > 1 \times 10^{16}\,\rm{GeV}$, are shown in green color.
We obtain that the upper bounds on the masses of 
both the first-two generation squarks and gluino are around $5~\rm {TeV}$,
and thus they are 
well within the reaches of the ${\rm FCC}_{\rm hh}$ and SppC~\cite{Cohen:2013xda,Golling:2016gvc}.  
Moreover, the neutralinos, charginos, and sleptons can be discovered
at the ${\rm FCC}_{\rm hh}$ and SppC as well.

\textbf{Gauge Mediated Supersymmetry Breaking.--}Finally, we study the 
Gauge Mediated Supersymmetry Breaking (GMSB)~\cite{Dine:1993yw, Giudice:1998bp,Meade:2008wd}. 
The GMSB is a method of communicating SUSY breaking to the SSMs from the hidden sector
 through the SM gauge interactions. The basic parameters of the minimal GMSB are:
   $\tan\beta$, ${\rm sign}(\mu)$,  the  messenger field mass scale $M_{mess}$,  
the number of $SU(5)$ representations of the messenger fields $N_{mess}$,  and 
the SUSY breaking scale in the visible sector $\Lambda$. The messenger fields induce 
the gaugino masses at one loop and then they are transmitted on to the squark and slepton masses at two loops.
To preserve the gauge coupling unification, we consider the messenger fields which form the complete
GUT multiplets. For simplicity, we introduce one pair of the messenger fields 
in the $\mathbf{5}$ and $\mathbf{\overline{5}}$ representations of $SU(5)$, {\it i.e.},
$N_{mess}=1$. Also, we take parameter $c_{grav}=1$.

We perform random scans over the following minimal GMSB parameter space
\begin{align}
5\times 10^{5} \, \rm{GeV}\,\leq  & \, \Lambda \,  \leq  10^{7}\, \rm{TeV}, \nonumber \\
2 \times \Lambda \, \leq  & \, M_{mess}    \leq  10^{15}\, \rm{GeV}, \nonumber  \\
2 \leq  & \, \tan\beta \, \leq  60
\label{parameterRange-GMSB}
\end{align}
with $\mu >0$ and $m_t = 173.2\, {\rm GeV}$ \cite{:2009ec}. Because $M_{GUT}$ is not calculated 
in all the current codes, 
we estimate $M_{GUT}$ indirectly by the following way. We take a benchmark point from the mSUGRA/CMSSM scenario 
with $M_{GUT}$ very close to $1\times 10^{16}\, \rm{GeV}$. 
With a specially modified version of the ISAJET, we define 
$\alpha_{12}^{-1}(Q)\equiv \alpha_{1}^{-1}(Q)-\alpha_{2}^{-1}(Q)$, and  
make a plot of $\alpha_{12}^{-1}(Q)$ from 
the Renormalization Group Equation (RGE) running of three guage couplings from $M_{GUT}$ to 
the weak scale $M_{W}$ as functions of renormalization scale $Q$.
 We then fit the $\alpha_{12}^{-1}(Q)$ curve by
 a polynomial function $f(Q)$ via Mathematica. For any point at the messenger scale $M_{mess}$, 
we calculate $\alpha_{1}^{-1}(M_{mess})$ and $\alpha_{2}^{-1}(M_{mess})$ via the codes SARAH 4.14.3 and Sphenov4.0.4. 
We use Spheno to do these calculations since it can compute and output the SM gauge couplings 
at the $M_{mess}$. Moreover, 
for $\alpha_{1}^{-1}(M_{mess})-\alpha_{2}^{-1}(M_{mess}) > f(M_{mess})$ and 
$\alpha_{1}^{-1}(M_{mess})-\alpha_{2}^{-1}(M_{mess}) < f(M_{mess})$, we obtain 
$M_{GUT} > 1 \times 10^{16}\,\rm{GeV}$ and $M_{GUT} < 1 \times 10^{16}\,\rm{GeV}$, respectively.
 Similarly, all the points, which satisfy the current experimental constraints and have
 $M_{GUT} > 1 \times 10^{16}\,\rm{GeV}$, are shown in green color 
in the right panel of Fig.~\ref{fig:GMSB}. We see that the upper bounds on the masses of 
the first-two generation squarks and gluino are $8\,\rm{TeV}$, and $6\,\rm{TeV}$, respectively.
Therefore, the first-two generation squarks and gluino are 
well within the reaches of the ${\rm FCC}_{\rm hh}$ and SppC~\cite{Cohen:2013xda,Golling:2016gvc}.  
Moreover, the neutralinos, charginos, and sleptons might be discovered
at the ${\rm FCC}_{\rm hh}$ and SppC as well.

\textbf{Summary and Conclusions.--}Gauge coupling unification in the SSMs strong suggests the GUTs.
Considering the grand desert hypothesis from the EW scale to GUT scale, 
we showed that the supersymmetric GUTs
can be probed at the future pp colliders and Hyper-Kamiokande experiment.
For the GUTs with $M_{GUT} \le 1.0\times 10^{16}$~GeV,
 the dimension-six proton decay via heavy gauge boson exchange can be probed at 
 the Hyper-Kamiokande experiment.
Moreover, for the GUTs with $M_{GUT} \ge 1.0\times 10^{16}$~GeV, we for the first time studied the
upper bounds on the gaugino and sfermion masses.
We showed that the supersymmetric GUTs with anomaly and gauge mediated supersymmetry
breakings are well within the reaches of the future 100 TeV pp colliders 
such as the ${\rm FCC}_{\rm hh}$ and SppC,
and the supersymmetric GUTs with gravity mediated supersymmetry breaking can be probed 
at the future 160 TeV pp collider. 
The interesting viable parameter spaces for gravity mediation,
which can be probed at the ${\rm FCC}_{\rm hh}$ and SppC, have been discussed as well.

\textbf{Acknowledgments.--}We would like to thank Jinmian Li and Qi-Shu Yan for helpful discussions. SR is also like to thank Howard Bare for useful discussion. This research was supported by 
the  Projects  11875062  and  11947302  supported  by  
the National  Natural  Science  Foundation  of  China,  
and  by the Key Research Program of Frontier Science, CAS.
The numerical results described in this paper have been obtained via the HPC Cluster of ITP-CAS, Beijing, China.%


.


\end{document}